\documentclass{aa}  

\usepackage{graphicx}
\usepackage{txfonts}
\usepackage{numprint}
\usepackage[squaren,Gray]{SIunits}
\usepackage{color}
\usepackage{amsmath}

\newcommand{\rlight}{r_{\rm L}}

\newcommand{\LL}{Landau-Lifshitz}
\newcommand{\RR}{radiation reaction}
\newcommand{\IRR}{improved \RR}

\begin{document} 

\title{A new radiation reaction approximation for particle dynamics in the strong field regime}

\author{J. P\'etri
}

\institute{Universit\'e de Strasbourg, CNRS, Observatoire astronomique de Strasbourg, UMR 7550, F-67000 Strasbourg, France.\\
\email{jerome.petri@astro.unistra.fr}         
}

\date{Received ; accepted }

 
  \abstract
   {Following particle trajectories in the intense electromagnetic field of a neutron star is prohibited by the large ratio between the cyclotron frequency~$\omega_{\rm B}$ and the stellar rotation frequency~$\Omega$. No fully kinetic simulations on a macroscopic scale and with realistic field strengths have been performed so far due to the huge computational cost implied by this enormous scale of separation.}
   {In this paper, we derive new expressions for the particle velocity subject to strong radiation reaction that are intended to be more accurate than the current state-of-the-art expression in the radiation reaction limit regime, the so-called Aristotelian regime.}
   {We shortened the timescale hierarchy by solving the particle equation of motion in the radiation reaction regime, where the Lorentz force is always and immediately balanced by the radiative drag, and including a friction not necessarily opposite to the velocity vector, as derived in the \LL\  approximation.}
   {Starting from the reduced \LL\ equation (i.e. neglecting the field time derivatives), we found expressions for the velocity depending only on the local electromagnetic field configuration and on a new parameter related to the field strength that controls the strength of the radiative damping. As an example, we imposed a constant Lorentz factor~$\gamma$ during the particle motion. We found that for ultra-relativistic velocities satisfying $\gamma \gtrsim 10$, the difference between strong radiation reaction and the \RR\ limit becomes negligible.}
   {The new velocity expressions produce results similar in accuracy to the radiation reaction limit approximation. We therefore do not expect this new method to improve the accuracy of neutron star magnetosphere simulations. The radiation reaction limit is a simple but accurate, robust, and efficient way to follow ultra-relativistic particles in a strong electromagnetic field.}


   \maketitle
%

\section{Introduction}

With the recent increase in computational power, performing full kinetic simulations of neutron star magnetospheres can now be envisaged. However, the difference in timescales between gyro frequency and stellar frequency prevents realistic values from being applied to these parameters. So far the only way to circumvent this scaling problem is to downsize these frequencies while still keeping them well separated by respecting the ordering of these frequencies. Although this is helpful for understanding the dynamics of charged particles in extreme environments, it does not permit estimations of the true efficiency of particle acceleration and radiation reaction to be made because the Lorentz factors reached are several orders of magnitude lower than the ones predicted from observations of very high energy photons (see the review by \cite{philippov_pulsar_2022}). Recently some attempts have been made to simulate realistic parameters by  \cite{tomczak_particle_2020} and \cite{petri_particle_2022-2}, but the computational time remains prohibitive, and only test particles have been investigated while neglecting their back reaction to the field.

It is highly desirable to overcome this limitation by employing an approximation known as the radiation reaction limit (RRL) regime, sometimes also called Aristotelian dynamics, for which the equation of motion with radiative friction is shortened by use of an algebraic expression for the particle velocity depending only on the local value of the electric and magnetic field. This idea was applied by \cite{mestel_axisymmetric_1985} and \cite{finkbeiner_effects_1989}. Spectra and light curves in this regime were extensively studied by \cite{petri_pulsar_2019} in a vacuum field. He found realistic Lorentz factors and photon energies in reasonable agreement with the spectra observed by Fermi/LAT \citep{abdo_second_2013}.

Recently, \cite{chang_trajectories_2022} generalised the RRL velocity by including the \LL\ term proportional to the velocity \citep{landau_physique_1989} and by computing the associated radiation spectra. They found a complicated formula that unfortunately does not apply to any electromagnetic field configurations. Moreover they introduced some hypotheses that are not well justified to derive an expression for the velocity. Following a different approach, \cite{cai_dynamics_2022} studied the validity of the RRL equilibrium by describing the particle motion in a Frenet frame with a finite Lorentz factor. They introduced the principal null directions, which are the eigenvectors of the electromagnetic field tensors. The spatial part is equal to the Aristotelian spatial velocity, or stated differently, it is equal to the RRL velocity. Although their analysis is based on the \LL\ equations, including the time evolution of the Lorentz factor, at the end of their derivation they had to resort to the computation of the curvature radius in order to estimate this aspect of the Lorentz factor. In this work, we attempt to estimate the Lorentz factor by evolving it in time from the initial conditions, but as we show, the curvature radiation interpretation leads to more accurate estimates of the Lorentz factor. \cite{cai_dynamics_2022} applied their idea to a rather artificial magnetic field configuration. Our aim is to apply such techniques to realistic fields, such as a rotating magnetic dipole.

In this paper, we derive formulas for the velocity in an arbitrary electromagnetic field configuration starting from the reduced \LL\ equation (LLR; i.e. neglecting the field time derivatives). 
In section~\ref{sec:Modele}, we derive the algorithm for the new velocity field according to the LLRs and that we call \IRR\ (IRR). Some explicit expressions of this velocity are given in Section~\ref{sec:Vitesse}. In Section~\ref{sec:Simulations}, we then quantify the improvement brought by the inclusion of the radiative friction term proportional to the velocity compared to the standard RRL. Conclusions and perspectives are touched on in Section~\ref{sec:Conclusions}.

\section{Strong radiation reaction regime}
\label{sec:Modele}

Radiation reaction can be thought of as a friction drag opposing some resistance to the Lorentz force. It acts as a brake and is appropriately depicted by a force opposite to the velocity vector. However, in the \LL\ approximation, the radiation reaction force is opposite to the velocity only in the limit of ultra-relativistic particles. For an arbitrary particle speed, there are additional components along the electric field $\vec{E}$, the magnetic field $\vec{B}$, and the electric drift motion $\vec{E} \wedge \vec{B}$. We aim to quantify the effect of these additional forces in the particle trajectory by first deriving a new expression for the velocity.

\subsection{Equation of motion}

As an approximation of the Lorentz-Abraham-Dirac equation, we employ the Landau-Lifshitz expression according to \cite{landau_physique_1989} such that
\begin{subequations}
        \label{eq:LL}
        \begin{align}
        \frac{du^i}{d\tau} & = \frac{q}{m} \, F^{ik} \, u_k + \frac{q \, \tau_{\rm m}}{m} \, g^i ,\\
        g^i & =  u^\ell \partial_\ell F^{ik} \, u_k + 
        \frac{q}{m} \, \left( F^{ik} \, F_{k\ell} \, u^\ell + ( F^{\ell m} \, u_m ) \, ( F_{\ell k} \, u^k ) \, \frac{u^i}{c^2} \right), 
        \end{align}
\end{subequations}
with the typical timescale related to the particle classical radius crossing time
\begin{equation}\label{eq:tau_m}
\tau_{\rm m} = \frac{q^2}{6\,\pi\,\varepsilon_0\,m\,c^3},
\end{equation}
with $\tau$ being the proper time, $u^i=\gamma(c,\vec{v})$ as the 4-velocity, $q$ being  the particle charge, $m$  as the mass, $c$ as the speed of light, $\mathbf{E}$ and $\mathbf{B}$ as the electric and magnetic field, $\varepsilon_0$ as the vacuum permittivity, $\mathbf{v}$ as the particle velocity, and $F^{ik}$ as the electromagnetic tensor. 

To derive the velocity vector~$\vec{v}$ and the Lorentz factor~$\gamma$, it is judicious to switch to the 3+1~formalism by introducing the observer time $dt = \gamma \, d\tau$. Therefore,
\begin{subequations}
        \label{eq:LL3D}
        \begin{align}
        \frac{d\vec{p}}{dt} & = q \, \vec{F}_L + \gamma \, q\,\tau_m \, \left[ \frac{d\vec{E}}{dt} + \vec v \wedge \frac{d\vec{B}}{dt} \right] \\
        & + \frac{q^2 \, \tau_m}{m} \, \left[ \vec F_L \wedge \vec B + ( \vec \beta \cdot \vec E ) \, \vec E/c \right] + \frac{q^2 \, \tau_m}{m\,c^2}  \, \gamma^2 \, [ ( \vec\beta \cdot \vec E)^2 - \vec F_L^2 ] \, \vec v \nonumber \\
        \frac{d\gamma}{dt} & = \frac{q}{mc} \, \left[ \vec\beta \cdot \vec E + \tau_m \, \gamma \, \vec{\beta} \cdot \frac{d\vec E}{dt} +\frac{q\,\tau_m}{m\,c} \, \left( \vec {F}_L \cdot \vec E + \gamma^2 \, [ ( \vec\beta \cdot \vec E)^2 - \vec F_L^2 ]  \right) \right],    
        \end{align}
\end{subequations}
where we define the vector field
\begin{equation}
\vec{F}_L = \vec{E} + \vec{v} \wedge \vec{B}
,\end{equation}
the normalised velocity $\vec{\beta} = \vec{v}/c$, and the momentum by $\vec{p} = \gamma\,m\,\vec{v}$.
In the constant field approximation, we drop the time derivatives and obtain the fundamental equation of motion for a particle as follows
\begin{subequations}
        \begin{align}
        \label{eq:mouvement3+1a}
        \frac{d\vec{p}}{dt} & = q \, \vec{F}_L + \frac{q^2 \, \tau_m}{m} \, \left[ \vec F_L \wedge \vec B + ( \vec \beta \cdot \vec E ) \, \vec E/c \right] \\
        & + \frac{q^2 \, \tau_m}{m\,c^2}  \, \gamma^2 \, [ ( \vec\beta \cdot \vec E)^2 - \vec F_L^2 ] \, \vec v \nonumber \\
        \label{eq:mouvement3+1b}
        \frac{d\gamma}{dt} & = \frac{q}{mc} \, \left[ \vec\beta \cdot \vec E +\frac{q\,\tau_m}{m\,c} \, \left( \vec {F}_L \cdot \vec E + \gamma^2 \, [ ( \vec\beta \cdot \vec E)^2 - \vec F_L^2 ]  \right) \right]   .
        \end{align}
\end{subequations}

\subsection{Derivation of the velocity: First approach}

The derivation of the particle velocity follows the procedure outlined by \cite{mestel_stellar_1999}. Nevertheless, instead of using a friction of the form $-K\,\mathbf{v}$ with $K>0$, we use the three-dimensional version of the radiation reaction force, neglecting the space-time dependence of the electromagnetic field such that the radiative force reduces to the second and third term in the right-hand side of Eq.~\eqref{eq:mouvement3+1a}.
Writing the radiation reaction force as
\begin{equation}\label{eq:f_rad_2}
\vec{F}^{\rm rad} = K_2 \, \left[ \vec{F}_{\rm L} \wedge \vec{B} + (\vec{\beta} \cdot \vec{E}) \, \vec{E}/c \right] - K_1 \, \vec{v}
\end{equation}
and balancing the Lorentz force
\begin{equation}\label{eq:f_ext}
\vec{F}^{\rm ext} = q \, \vec{F}_{\rm L} 
\end{equation}
with this radiation reaction $\vec{F}^{\rm ext} = \vec{F}^{\rm rad}$, we arrive at
\begin{equation}
\label{eq:AristoteGenerale}
q \, \vec{F}_{\rm L} = ( K_1 + K_2 \, B^2 ) \, \vec v - K_2 ( \vec{E} \wedge \vec{B} + (\vec{B} \cdot \vec{v}) \, \vec{B} + (\vec{\beta} \cdot \vec{E}) \, \vec{E}/c ) .
\end{equation}
We note that there is no assumption about particles moving at the speed of light, their Lorentz factor is arbitrary, and $v<c$. This represents a novelty compared to all other radiation reaction expressions, which are always enforcing $v=c$.
The coefficients $K_1$ and $K_2$ are deduced from Eq.~\eqref{eq:mouvement3+1a} and given by
\begin{subequations}
        \label{eq:K_12}
        \begin{align}
        K_1 & = \frac{q^2 \, \tau_m}{m\,c^2} \, \gamma^2 \, \left[ \vec{F}_{\rm L}^2 - (\vec{\beta} \cdot \vec{E})^2 \right] \\
        K_2 & = \frac{q^2 \, \tau_m}{m} .
        \end{align}
\end{subequations}
We note that these coefficients are algebraic, being positive whatever the sign of the charge~$q$.
However, $K_1$ depends on the Lorentz factor~$\gamma$, and we leave it unconstrained. In order to solve Eq.~\eqref{eq:AristoteGenerale}, the velocity is advantageously decomposed into three components $(\sigma, \delta, \eta)$ such that 
\begin{equation}\label{eq:VRR}
\vec{v} = \sigma \, \vec{E} + \delta \, \vec{B} + \eta \, \vec{E} \wedge \vec{B} .
\end{equation}
These components must satisfy a linear system of three equations of unknowns $(\sigma, \delta, \eta)$ according to
\begin{subequations}\label{eq:systeme}
        \begin{align}
        q \, (1 - \eta \, B^2) & = [K_1 + K_2 \, B^2 ] \, \sigma - K_2 \, [ \sigma \, {E}^2 + \delta \, (\vec{E} \cdot \vec{B}) ] / c^2 \\
        q \, \eta \, (\vec{E} \cdot \vec{B}) & = K_1 \, \delta - K_2 \, (\vec{E} \cdot \vec{B}) \, \sigma \\
        q \, \sigma & = [ K_1 + K_2 \, B^2 ] \, \eta - K_2 .
        \end{align}
\end{subequations}
We recall that $K_1$ is unconstrained. Therefore, in order to fully solve the system, an additional condition is required for $K_1$. To this end, we could enforce $v=c$, but as a generalisation, we impose an user-defined Lorentz factor $\gamma = (1-v^2/c^2)^{-1/2}$.
Equations~\eqref{eq:VRR} and \eqref{eq:systeme}, supplemented with the condition on the speed $v$, fully determine the velocity vector $\mathbf{v}$. Solving for $\delta$, we get
\begin{equation}\label{eq:delta}
 \delta = \frac{\vec{E} \cdot \vec{B}}{K_1} \, [ q \, \eta + K_2 \, \sigma ],
\end{equation}
reducing the system to a 2x2 size. Indeed, the smaller linear system to be solved reads
\begin{equation}\label{eq:K1pK2}
	\small
\begin{pmatrix}
K_1 + K_2 \, \left( B^2- \frac{E^2}{c^2} \right) - \frac{K_2^2}{K_1} \, \left(\frac{\vec{E} \cdot \vec{B}}{c} \right)^2 & q \, \left[ B^2 - \frac{K_2}{K_1} \, \left(\frac{\vec{E} \cdot \vec{B}}{c} \right)^2 \right] \\
-q & K_1 + K_2 \, B^2 
\end{pmatrix}
\begin{pmatrix}
\sigma \\ \eta
\end{pmatrix}
=
\begin{pmatrix}
q \\ K_2
\end{pmatrix}.
\end{equation}
Deviation from the standard RRL arises because of the terms containing $K_2$, which are usually neglected in the ultra-relativistic regime. 
In order to deduce the number of relevant free parameters in the problem, it is preferable to employ quantities without dimensions, as explained in the next section.

\subsection{Dimensionless system}

As generally required for numerical simulations, we introduce several useful quantities without dimensions relevant for the computation of the velocity. Following our previous work in \cite{petri_particle_2022-2}, the primary fundamental variables are: the speed of light~$c$; a typical frequency $\omega$ involved in the problem; the particle electric charge $q$; and the particle rest mass $m$.
From these quantities we derive a typical time and length scale as well as electromagnetic field strengths such that the length scale $L = c/\omega$;
the time scale $T = 1/\omega$; the magnetic field strength~$B_n = m\,\omega/|q|$; the electric field strength~$E_n = c\,B_n$; and the typical electromagnetic force strength~$F_n = |q| \, E_n$.
The two important parameters defining the family of solutions are the field strength parameters $a_{B}$ and $a_{E}$ and the radiation reaction efficiency $k_2 = \omega\,\tau_{\rm m}$ according to the following definitions
\begin{subequations}
        \label{eq:Parametres}
        \begin{align}
        a_{B} & = \frac{B}{B_n} = \frac{\omega_{\rm B}}{\omega} \\
        a_{E} & = \frac{E}{E_n} = \frac{\omega_{\rm E}}{\omega} .
        \end{align}
\end{subequations}
The external force becomes
\begin{equation}\label{eq:f_ext_norm}
 \frac{\vec{F}^{\rm ext}}{F_n} = \textrm{sign}(q) \, ( \vec{e} + \vec{\beta} \wedge \vec{b} )
\end{equation}
with $\textrm{sign}(q)=q/|q|$ and the radiative force
\begin{equation}\label{eq:f_rad_2_norm}
\frac{\vec{F}^{\rm rad}}{F_n} = k_2 \, \left[ \vec{e} \wedge \vec{b} + \vec{b} \wedge ( \vec{b} \wedge \vec{\beta} ) +  (\vec{\beta} \cdot \vec{e}) \, \vec{e} \right]
- k_1 \, \vec{\beta}
\end{equation}
with the normalised fields $\vec{e} = \vec{E}/E_n$,  $\vec{b} = \vec{B}/B_n$ and 
\begin{equation}\label{eq:k_1}
k_1 = \frac{K_1}{|q|\,B_n} \qquad ; \qquad k_2 = \frac{K_2 \, B_n}{|q|} = \omega \, \tau_{\rm m} .
\end{equation}
The velocity expansion coefficients are also normalised according to
\begin{equation}
\tilde{\sigma} = \sigma \, B_n \qquad ; \qquad \tilde{\eta} = \eta \, B_n^2 \qquad ; \qquad \tilde{\delta} = \delta \, B_n/c.
\end{equation}
The normalised system to be solved then reads with $\zeta=sign(q)$
\begin{equation}\label{eq:systeme_2}
	\small
\begin{pmatrix}
k_1 + k_2 \, \left( b^2- e^2 \right) - \frac{k_2^2}{k_1} \, \left( \vec{e} \cdot \vec{b} \right)^2 & \zeta\left[b^2 - \frac{k_2}{k_1} \, \left(\vec{e} \cdot \vec{b} \right)^2\right] \\
-\zeta & k_1 + k_2 \, b^2 
\end{pmatrix}
\begin{pmatrix}
\tilde{\sigma} \\ \tilde{\eta}
\end{pmatrix}
=
\begin{pmatrix}
\zeta \\ k_2
\end{pmatrix}
.
\end{equation}
In the above system, $k_2$ is fixed by the nature of the charged particle ($q,m$) and the typical frequency~$\omega$. There is no freedom to choose it arbitrarily. However $k_1$ is undetermined and needs to be fixed by an additional constraint on the velocity. Choosing the Lorentz factor~$\gamma$, the coefficient $k_1$ is found from the condition $\lVert \mathbf{v} \rVert / c = 1 - \gamma^{-2}$.

Actually $k_1$ is related to the velocity~$\vec{v}$ by Eq.~\eqref{eq:K_12} in the LLR approximation. But solving for the coefficients $\sigma,\delta,\eta$, and $\vec{v}$ is only a function of $k_1$. Thus Eq.~\eqref{eq:K_12} is a non-linear equation for $k_1$ solely, which connects back to the fact that Eq.~\eqref{eq:AristoteGenerale} is a non-linear equation for $\vec{v}$ involving the Lorentz factor.
This first approach has the drawback of implicitly including the Lorentz factor in the linear system via the parameter $K_1$. In the next sub-section, we develop a second approach that is quadratic in the velocity and does not contain the Lorentz factor.

\subsection{Derivation of the velocity: Second approach}

In a second approach, called velocity radiation reaction (VRR), instead of cancelling the relativistic momentum time derivative $\frac{d\vec{p}}{dt}$, we decided to cancel the velocity time derivative $\frac{d\vec{v}}{dt}$ given in the \LL\ approximation by
\begin{multline}\label{eq:vitesse2emeapproche}
\gamma \, m \, \frac{d\vec{v}}{dt} = q [\vec E + \vec v \wedge \vec B - (\vec{\beta} \cdot \vec{E}) \vec{\beta}] + \\
\frac{q^2 \, \tau_{\rm m}}{m} \left[ \vec E \wedge \vec B + (\vec v \wedge \vec B) \wedge \vec B + (\vec \beta \wedge \vec E) \wedge \vec E / c + \vec \beta \cdot (\vec E \wedge \vec B) \, \vec \beta \right] .
\end{multline}
The advantage of this approach is that it sticks closer to the Aristotelian regime. Indeed, if the term involving $\tau_{\rm m}$ is removed, we retrieve the RRL and the associated Aristotelian velocity expression that exactly satisfies
\begin{equation}\label{eq:AristoteExacte}
\vec E + \vec v \wedge \vec B - (\vec{\beta} \cdot \vec{E}) \vec{\beta} = 0.
\end{equation}
Translated into normalised units, we get
\begin{multline}\label{eq:root}
\zeta \, k_2 \, \left[ \vec e \wedge \vec b + (\vec \beta \wedge \vec b) \wedge \vec b + (\vec \beta \wedge \vec e) \wedge \vec e + \vec \beta \cdot (\vec e \wedge \vec b) \, \vec \beta \right] \\
+ \vec e + \vec \beta \wedge \vec b - (\vec{\beta} \cdot \vec{e}) \vec{\beta} = 0.
\end{multline}
This expression is quadratic in $\vec{\beta,}$ and unlike the previous approach, it does not involve the Lorentz factor. It can thus be solved by standard root finding techniques for a fixed Lorentz factor (or equivalently a fixed velocity norm). In a simple prescription, we set the velocity norm to $\|\vec{v}\| = c$ again, but any Lorentz factor can be imposed. Departure from the RRL arises due to the term involving $\zeta \, k_2$. In the next section, we discuss the different approximations to the particle Lorentz factor.

\section{Approximations of the particle Lorentz factor}
\label{sec:Vitesse}

In this section, we explore several approximations to estimate the particle velocity without resorting to a full time integration of the equation of motion. We first remind the standard expression for the velocity, how it compares to our new expression and then discuss the Lorentz factor estimation.

\subsection{Friction opposite to velocity}

Starting from the radiation reaction description of \cite{mestel_stellar_1999} where the radiative friction is opposite to the particle velocity vector~$\vec{v}$, we write
\begin{equation}\label{eq:RRL}
q \, (\vec E + \vec v \wedge \vec B) = K \, \vec{v},
\end{equation}
where $K$ is a positive parameter related to the power radiated by the particle.
Solving for the velocity, we find
\begin{equation}\label{eq:VRRL_K}
\left( B^2 + \frac{K^2}{q^2} \right) \, \vec{v} = \frac{K}{q} \, \vec{E} + \vec E \wedge \vec B + \frac{q}{K} \, ( \vec{E} \cdot \vec{B} ) \vec{B}.
\end{equation}
Moreover $K$ is the only positive solution of the bi-quadratic equation
\begin{equation}
K^4 \, v^2 - q^2 \, ( E^2 - v^2 \, B^2 ) \, K^2 - q^4 \, (\vec E \cdot \vec B)^2 = 0 .
\end{equation}
Thus, it satisfies
\begin{equation}\label{eq:Ksolution}
\frac{K}{|q|} = \sqrt{ \frac{E^2 - v^2 \, B^2 + \sqrt{( E^2 - v^2 \, B^2 )^2 + 4 \,v^2 \,(\vec{E} \cdot \vec{B})^2}}{2\,v^2}} .
\end{equation}
So far there are no constraints on the particle speed $v<c$. In the limit of $v=c$, we retrieve the velocity expression used in the literature, namely,
\begin{equation}
        \label{eq:VRRL}
        \mathbf{v}_\pm = \frac{\mathbf{E} \wedge \mathbf{B} \pm ( E_0 \, \mathbf{E} / c + c \, B_0 \, \mathbf{B})}{E_0^2/c^2+B^2},
\end{equation}
assuming particles moving at the speed of light $\lVert\mathbf{v}_\pm\rVert=c$. In the equation, $\vec{v}_+$ represents positively charged particles, whereas $\vec{v}_-$ represents negatively charged particles. The electromagnetic field strength $E_0$ and $B_0$ are deduced from the electromagnetic invariants $\mathbf E^2 - c^2 \, \mathbf B^2 = E_0^2 - c^2 \, B_0^2$ and $\mathbf E \cdot \mathbf B = E_0 \, B_0$ with the constraint $E_0\geq0$. Therefore, the radiated power is $\mathcal{P}_R = q\,\vec{F}_L \cdot \vec{v} = |q| \, c \, E_0$ and $K = |q| \, E_0 / c \geq0$.

Except for this ultra-relativistic limit for which we assume $v=c$, another relation is required to set the particle Lorentz factor~$\gamma$. To this end, we equate the radiated power according to the local curvature radius~$\rho_c$ of the particle trajectory as
\begin{equation}\label{eq:puissance_courbure}
        \mathcal{P}_R = \frac{q^2}{6\,\pi\,\varepsilon_0} \, \gamma^4 \, \frac{c}{\rho_c^2} = \gamma^4  \frac{\tau_{\rm m} \, m\,c^4}{\rho_c^2} = |q| \, c \, E_0
\end{equation}
from which the Lorentz factor becomes
\begin{equation}\label{eq:gamma_courbure}
\gamma = \left( \frac{|q|\,E_0}{\tau_{\rm m} \, m \, c^3} \, \rho_c^2\right)^{1/4} = \left( \frac{\vec{\beta} \cdot \vec{e}}{k_2} \, \frac{\rho_c^2}{\rlight^2}\right)^{1/4} .
\end{equation}
Moreover, the curvature is found from the acceleration by
\begin{equation}\label{eq:kappa_c}
\kappa_c = \frac{1}{\rho_c} \approx \left\lVert \frac{d\vec{\beta}}{c\,dt} \right\rVert .
\end{equation}
As long as $\gamma \gg 1$, the RRL Eq.~\eqref{eq:VRRL} remains a very good approximation.

In Eq.~\eqref{eq:RRL}, the strength of the damping $K$ is undetermined but usually set by the particle velocity $v$ or equivalently by its Lorentz factor~$\gamma$. Looking at LLR, we observed that the radiation reaction force term proportional to $\gamma^2$ is also opposite to the velocity~$\vec{v}$. We could therefore identify $K_1$ with $K$ to get
\begin{equation}\label{eq:K_from_LLR}
K \, \tau_{\rm m} \, v^2 = m \, c^2 .
\end{equation}
Hence, $K/|q|\geq m / |q| \, \tau_{\rm m} = 9\times 10^{11}$~T for electrons and positrons. This value is much too high.
As a consequence, there are two equations, Eqs.~\eqref{eq:Ksolution} and \eqref{eq:K_from_LLR}, for the two unknowns $K$ and $v$. 

In the ultra-relativistic limit $K\approx |q|\,E_0/c$ and 
\begin{equation}
 \beta^2 \approx \frac{m\,c}{|q|\,E_0\,\tau_{\rm m}} = \frac{1}{\omega_{\rm E_0} \, \tau_{\rm m}}.
\end{equation}
Keeping the velocity less than the speed of light leads to $E_0 \geq 2.7\times10^{20}$~V/m, which is even larger than the critical value of $E_{\rm crit} \approx 1.3\times10^{18}$~V/m. Therefore, this idea fails and gives the same value as before for the magnetic equivalent of $c\,E_0 = 9\times 10^{11}$~T.
We must conclude that the only reasonable way to compute the Lorentz factor is via the curvature radiation power Eq.~\eqref{eq:puissance_courbure}.
Before switching back to the LLR equation, we check how the new velocity approximation compares to the simple prescription presented in this paragraph.

\subsection{Comparison to the radiation reaction limit}

In the literature about approximated radiation reaction formulas, only the force opposite to the velocity is considered. Translated into our more general approach dealing with the full set of terms in the LLR equation, we enforce $k_2=0$. The system \eqref{eq:systeme_2} then simplifies into
\begin{equation}\label{eq:systeme_3}
\begin{pmatrix}
k_1 & \zeta \, b^2 \\
-\zeta & k_1
\end{pmatrix}
\begin{pmatrix}
\tilde{\sigma} \\ \tilde{\eta}
\end{pmatrix}
=
\begin{pmatrix}
\zeta \\ 0
\end{pmatrix} .
\end{equation}
The solution is readily found with
\begin{subequations}
        \begin{align}
                \tilde{\eta} & = \frac{1}{k_1^2 + b^2} \\
                \tilde{\sigma} & = \zeta \, k_1 \, \tilde{\eta} \\
                \tilde{\delta} & = \zeta \, \frac{\vec{e} \cdot \vec{b}}{k_1} \, \tilde{\eta} .
        \end{align}
\end{subequations}
This solution is exactly the same as the one presented in Eq.~\eqref{eq:VRRL_K} except that all quantities are now  normalised.
The speed is not explicitly imposed to be equal to the speed of light. Therefore, $k_1$ needs to be deduced, for instance, from the Lorentz factor, as in the previous discussion about curvature radiation power.

When $k_2\neq0$, $k_1$ is the root of a polynomial of high degree with no analytical expression. We compute this coefficient by applying a root finding algorithm via Newton-Raphson. A good initial guess for $k_1$ in the system~\eqref{eq:systeme_2} is 
\begin{equation}
 k_1 \approx \frac{E_0}{E_n} = \frac{|q|\,E_0}{m\,c\,\omega} = a_{E_0} .
\end{equation}
We checked that very few iterations are required to converge to a highly accurate solution with several digits of precision. Some simulations are shown in the next section. Finally, in the last approach, we use the full terms in the original LLR equation and solve for the velocity while taking into account the term independent of $\gamma$.

\subsection{Lorentz factor from LLR}

The system of equations~\eqref{eq:systeme_2} solves the particle velocity vector by assuming the coefficient $k_1$ is freely adjustable. Actually, from the LLR equation, it is not tuneable and must be determined self-consistently with the expression~\eqref{eq:K_12} in which there are no free parameters once the velocity $\vec{v}$ is fixed. 
This approach would lead to a first algorithm for finding the particle Lorentz factor~$\gamma$. We need to solve for $\gamma$ such that Eq.~\eqref{eq:K_12} is verified. However, as we show in the next section, the IRR algorithm finds very similar velocities compared to the `standard' algorithm. In this case Eq.~\eqref{eq:RRL} also holds, approximately. Then it can be shown that
\begin{equation}
\gamma^2 \, q^2 \, \left[ (\vec{\beta} \cdot \vec{E})^2 - \vec{F}_L^2 \right] \approx - K_1^2\,v^2,
\end{equation}
which gives values of $K_1$ very different from the expectation in Eq.~\eqref{eq:K_12}.

In a second alternative algorithm using the LLR equation, we can try to set $d\gamma/dt=0$ and solve for the value of $K_1$ such that it satisfies
\begin{equation}\label{eq:Lorentz_Limit}
\vec\beta \cdot \vec E +\frac{q\,\tau_m}{m\,c} \, \left( \vec {F}_L \cdot \vec E + \gamma^2 \, [ ( \vec\beta \cdot \vec E)^2 - \vec F_L^2 ]  \right) = 0.
\end{equation}
Here again there are no free parameters once the velocity $\vec{v}$ is fixed. 
This equation constrains the Lorentz factor because it depends on $\vec{v},$ which is fully solved once $K_1$ is fixed. Therefore, the procedure consists of finding the root of Eq.~\eqref{eq:Lorentz_Limit} depending only on $\gamma$. However, here also, as the solution is close to the expression~\eqref{eq:VRRL}, we found instead that
\begin{equation}\label{eq:zero}
 \vec {F}_L \cdot \vec E + \gamma^2 \, [ ( \vec\beta \cdot \vec E)^2 - \vec F_L^2 ]  \approx 0,
\end{equation}
which is not compatible with Eq.~\eqref{eq:Lorentz_Limit}.

Actually, the second term in Eq.~\eqref{eq:Lorentz_Limit} corresponds to the opposite of the curvature radiation power~$\mathcal{P}_R$. It is related to the curvature~$\kappa_c$ in the case of an ultra-relativistic particle such that
\begin{equation}
\kappa_c = \frac{|q|}{\gamma^2\,m\,c^2} \, \sqrt{\gamma^2 \, [ \vec F_L^2 - ( \vec\beta \cdot \vec E)^2 ] - \vec {F}_L \cdot \vec E } .
\end{equation}
If the term $\vec {F}_L \cdot \vec E$ is negligible, we retrieve the result \citep{kelner_synchro-curvature_2015}
\begin{equation}
\kappa_c \approx \frac{|q|}{\gamma\,m\,c^2} \, \sqrt{\vec F_L^2 - ( \vec\beta \cdot \vec E)^2} .
\end{equation}
The curvature would vanish in the limiting case investigated in this section because, by construction, $d\vec{v}/dt = \vec{0}$. Consequently, as in the previous section, the best procedure to compute the Lorentz factor is through the curvature radiation power~$\mathcal{P}_R$, again by replacing $E_0$ by $\vec\beta \cdot \vec E$ in Eq.~\eqref{eq:puissance_courbure} and Eq.~\eqref{eq:gamma_courbure}.

A final trial consisted of integrating the Lorentz factor differential equation~\eqref{eq:mouvement3+1b} in time from the initial conditions. Because the regime is close to the RRL, the second term expressing the power radiated almost always vanishes. Contrary to the magnetic field, only the electric field produces work and is able to accelerate particles. The results are less good compared to the curvature radius approach. Actually, the curvature radius represents only an auxiliary variable to compute the Lorentz factor. It could be derived straightforwardly from the definition of Eq.~\eqref{eq:kappa_c} but at the expense of computing the Lagrangian time derivatives of the electric and magnetic fields as
\begin{equation}\label{key}
        \frac{d\vec{E}}{dt} = \frac{\partial\vec{E}}{\partial t} + \vec{v} \cdot \vec{\nabla} \vec E
\end{equation}
and with a similar expression for $\vec{B}$.
These expressions are, however, unwieldy to implement because they require the computing of partial time and space derivatives $\partial_t$ and $\partial_{\vec{r}}$. We prefer to compute the curvature from a finite difference approximation of Eq.~\eqref{eq:kappa_c}, which is an equivalent description but much simpler to implement numerically.
In the next section, we explore the efficiency and accuracy of the above mentioned methods for a rotating magnetic dipole with an electric quadrupole component.

\section{Simulations around a rotating dipole}
\label{sec:Simulations}

As a typical macroscopic frequency, we used the neutron star rotation frequency~$\Omega$ and set $\omega = \Omega$. The numerical setup, electromagnetic field configuration, and initial conditions for particle position and velocity are exactly the same as in \cite{petri_particle_2022-2}. We simulated a sample of test particles evolving in the \cite{deutsch_electromagnetic_1955} electromagnetic field.

\subsection{Radiation reaction limit accuracy}

\begin{figure}
        \centering
        \includegraphics[width=\linewidth]{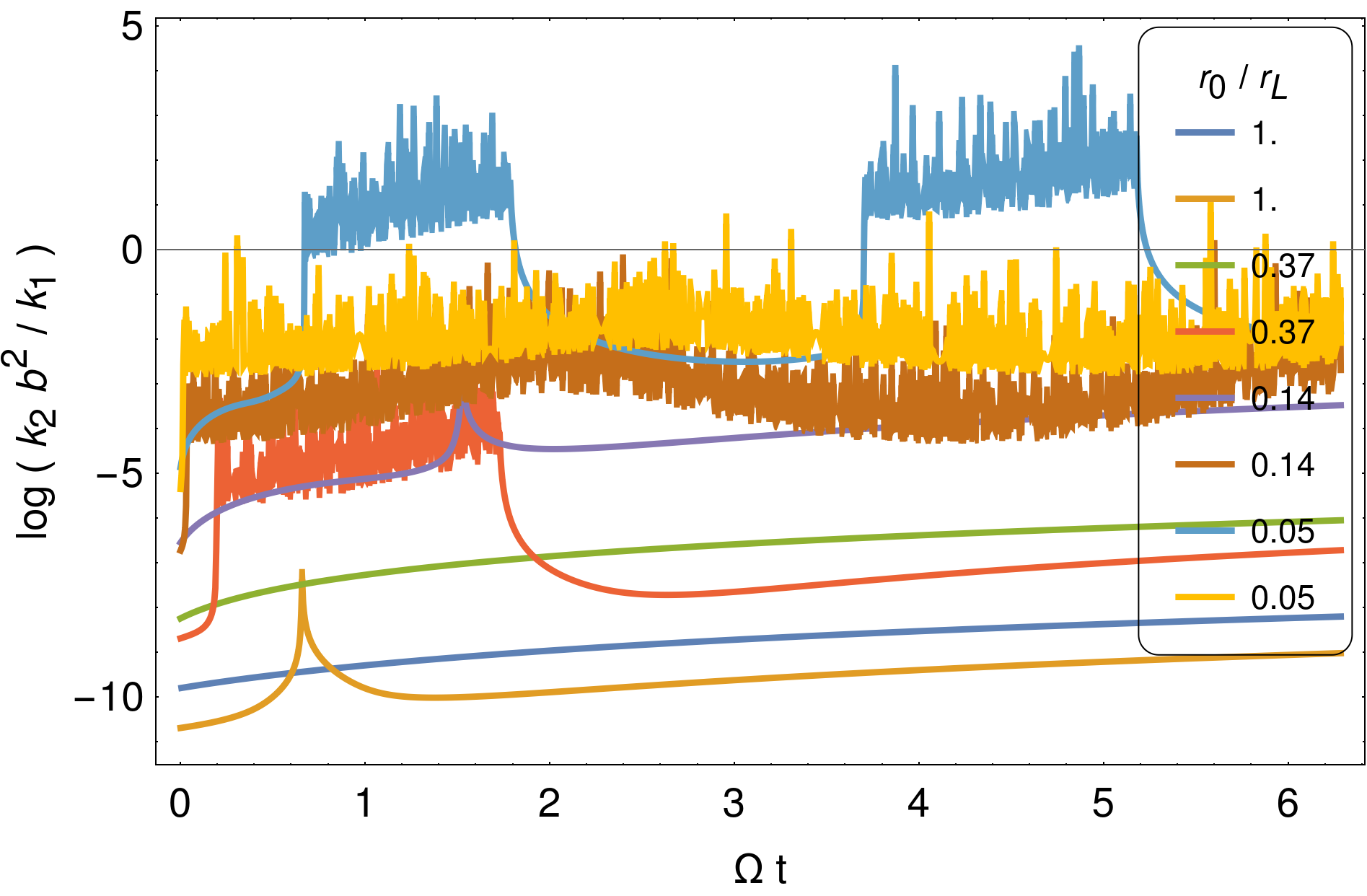}
        \caption{Ratio of the coefficient $k_1$ and $k_2$ expressed as $k_2\,b^2/k_1$ in log scale for a sample of eight~trajectories starting at several distances from the surface given by $r_0/\rlight \approx \{1, 0.37, 0.14, 0.05\}$. \label{fig:electronc60a13g10aramcfl-1k2b2sk1}}
\end{figure}

The improved version of the radiation reaction regime in the first approach differs significantly from the standard version only whenever the ratio $k_2\,b^2/k_1$ becomes comparable or greater than one. This means that the braking force no longer aligns with the particle velocity vector and that it also involves friction in the $\vec{E}$, $\vec{B}$, and $\vec{E} \wedge \vec{B}$ directions. To check if this situation happens for electrons in the rotating magnetic dipole, we plotted this ratio in a log scale (see Fig.\ref{fig:electronc60a13g10aramcfl-1k2b2sk1}) for a sample of eight~trajectories starting at different locations~$r_0$ within the light cylinder. The radial distances are given in the legend of the figure and were normalised to the light cylinder radius~$\rlight$ such that $r_0/\rlight \approx \{1, 0.37, 0.14, 0.05\}$. As can be seen in this plot, this ratio is mostly much lower than one, meaning that the improved version of radiation reaction does not significantly differ from the straightforward RRL, except for sparse events of very few trajectories. Moreover, as we later show, even in these cases, the trajectories are not drastically affected by the corrections brought through the IRR expression.

\begin{figure}
        \centering
        \includegraphics[width=0.95\linewidth]{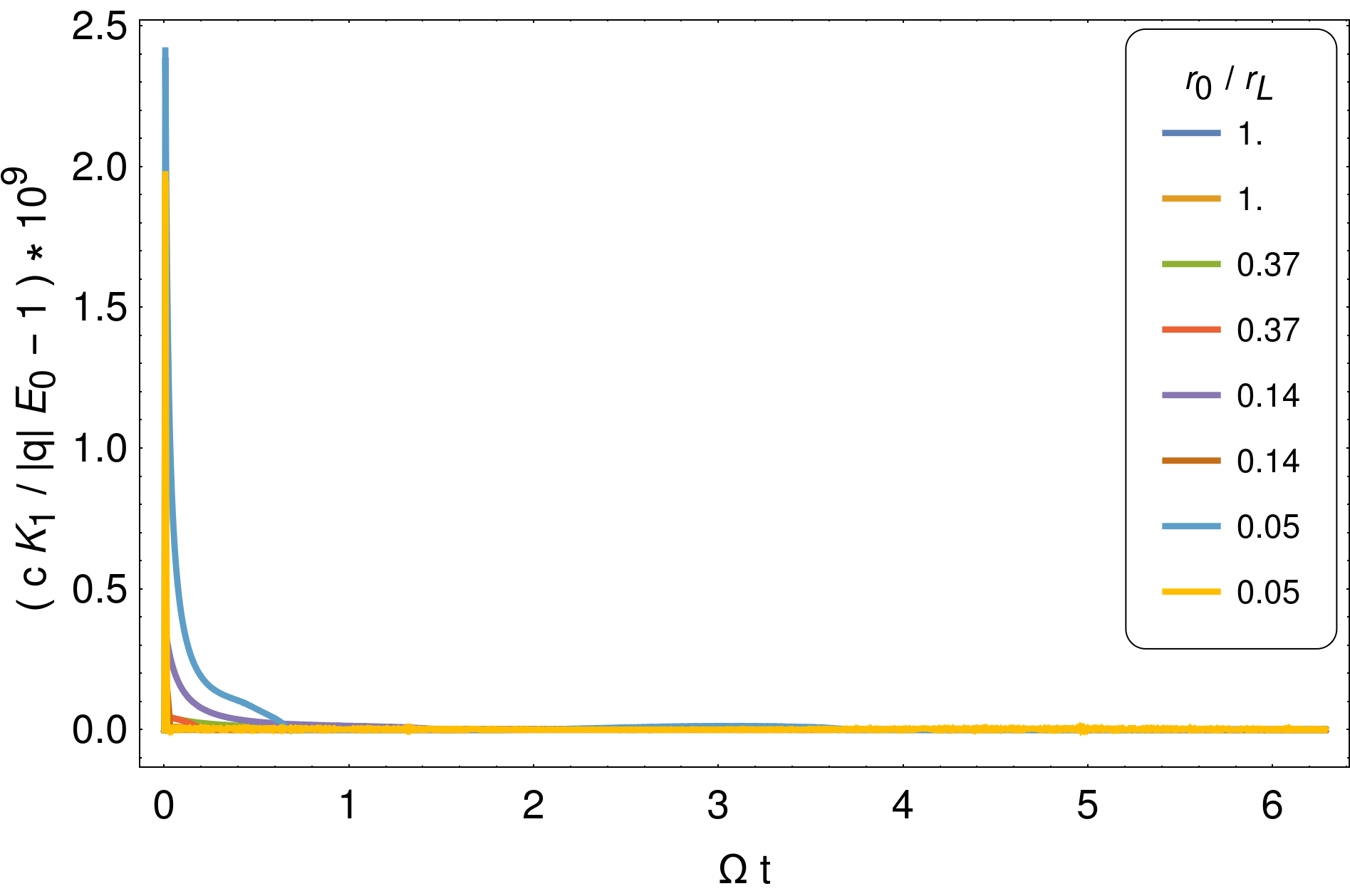}
        \caption{Time evolution of the ratio $c\,K_1/|q|\,E_0$ in log scale for a sample of eight representative trajectories. Both numbers are identical to eight digits of precision.}
        \label{fig:electronc60a13g6aramcfl-3k1}
\end{figure}

\begin{figure}
        \centering
        \includegraphics[width=0.95\linewidth]{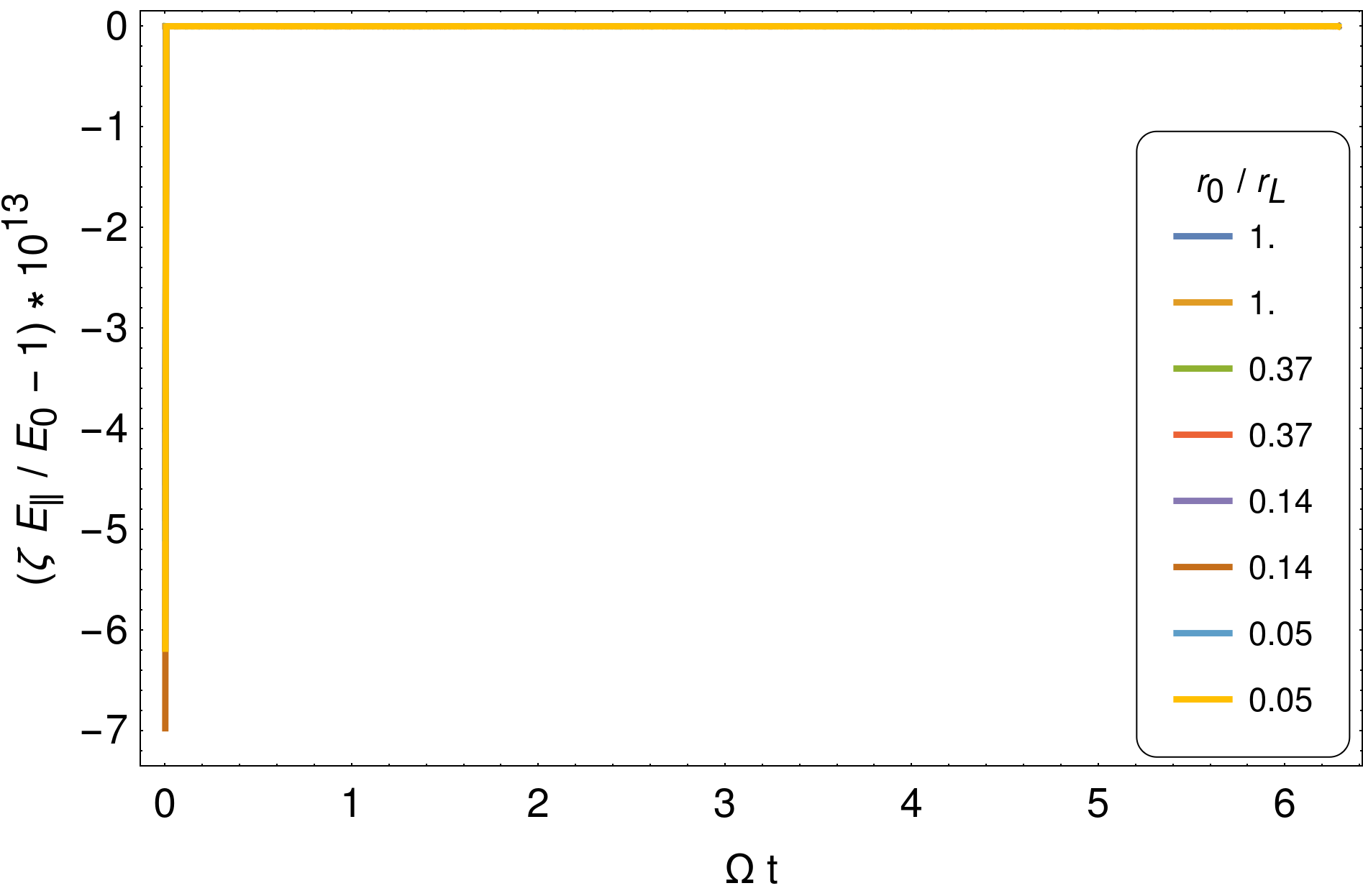}
        \caption{Time evolution of the true parallel electric field~$E_\parallel$ compared to the estimated parallel electric field~$E_0$ for a sample of eight trajectories. Because $E_\parallel<0$ is negative for electrons, we plot $\zeta \, E_\parallel/E_0$.}
        \label{fig:electronc60a13g6aramcfl-3eparrallel}
\end{figure}

Figure~\ref{fig:electronc60a13g6aramcfl-3k1} shows the deviation of $K_1$ from $|q|\,E_0/c$ in the IRR\ version for the same sample shown in Fig.~\ref{fig:electronc60a13g10aramcfl-1k2b2sk1}. The ratio equals one to very high accuracy. Both parameters are identical up to eight digits of precision. This supports the fact that the improvement is marginal.
Finally, in Fig.~\ref{fig:electronc60a13g6aramcfl-3eparrallel}, we compare the parallel electric field $E_\parallel = \vec{\beta} \cdot \vec{E}$ to the value~$E_0$ corresponding to the parallel electric field in the strict radiation reaction regime. Because $E_\parallel<0$ for negatively charged particles, we plotted $\zeta \, E_\parallel/E_0$ to keep positive numbers. Both values of the parallel electric field are identical to more that 12~digits of precision.

Based on all the above observations, we did not expect to observe a drastic change in the particle dynamics between the RRL and IRR regime. For a more quantitative analysis, we plotted the Lorentz factor evolution in time for the same sample of electrons. No difference in the Lorentz factor evaluation was observed between both approximations.
We therefore concluded that there is no advantage in including the correction brought by the \LL\  equation with a friction term not anti-aligned with the velocity vector.

\subsection{Comparison between IRR, VRR, and LLR}

We have checked that the IRR does not bring significant improvements compared to \RR. To complete the analysis of accuracy and efficiency of the IRR approximation, we compared it to the more reliable LLR equation of motion.

\begin{figure}
        \centering
        \includegraphics[width=0.95\linewidth]{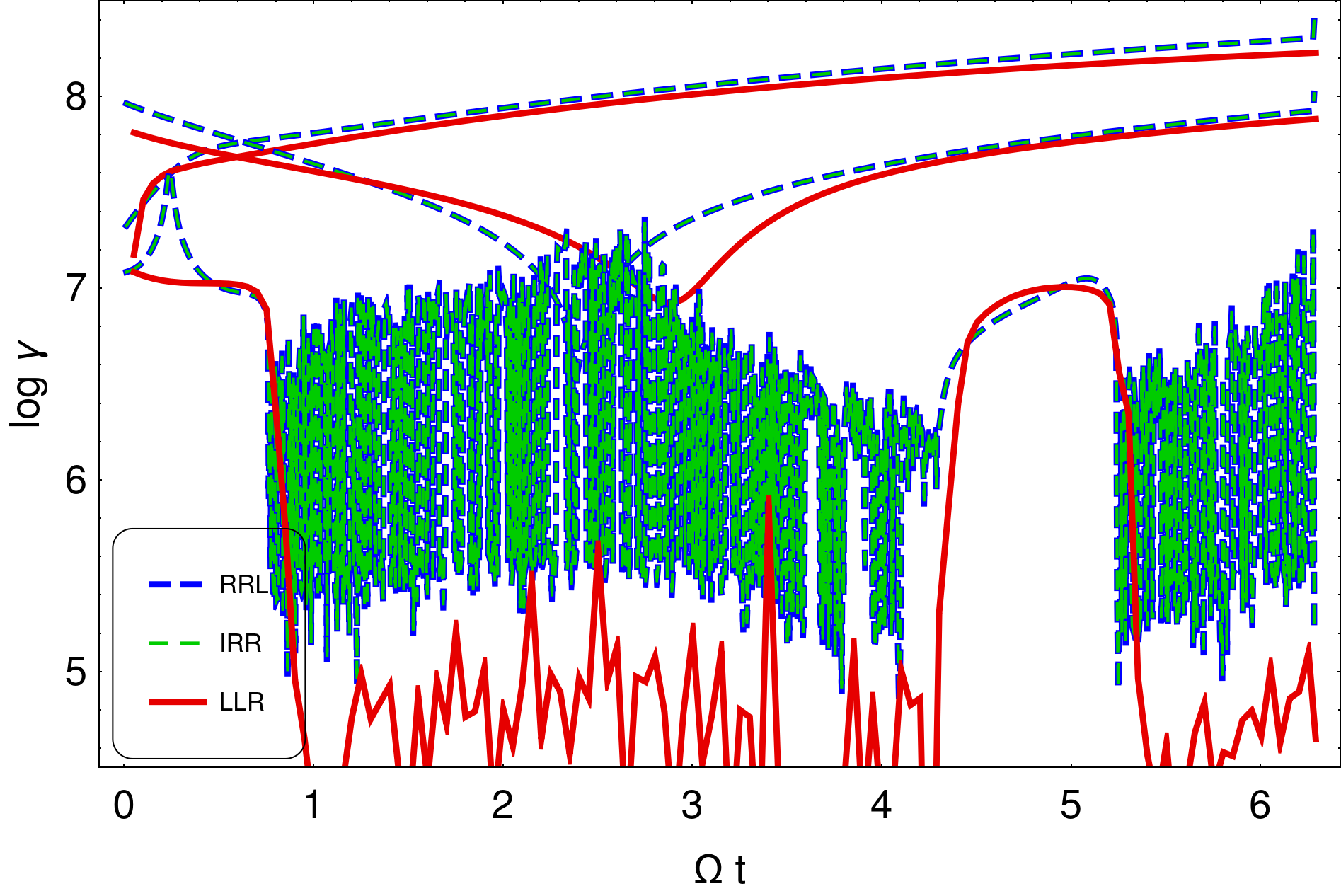}
        \caption{Evolution of the Lorentz factor for the three approximations of the equation of motion of an electron. Colours are as follows: RRL in dashed blue, IRR in dashed green, and LLR in solid red lines.}
        \label{fig:electronc60a13g6cfl-3gamma}
\end{figure}

Figure~\ref{fig:electronc60a13g6cfl-3gamma} shows the evolution of the Lorentz factor for the three descriptions of the particle motion, RRL, IRR, and LLR. The curves only differ by their initial condition. The RRL and its improved version show Lorentz factor estimates agreeing with the LLR computations to reasonably good accuracy. We note that the time evolution of the Lorentz factor is reproduced with the associated fluctuations for one of the trajectories. For the RRL and IRR case, the Lorentz factors were computed according to expression~\eqref{eq:gamma_courbure}. The curvature~$\kappa_c$ in Eq.~\eqref{eq:kappa_c} was estimated with a finite difference approximation
\begin{equation}\label{eq:kappa_approx}
        \kappa_c = \left\lVert \frac{\vec{v}^{n+1/2}-\vec{v}^{n-1/2}}{c^2 \, dt} \right\lVert .
\end{equation}
It only involved the value of the electromagnetic field at two neighbouring times: $t^{n+1/2}$ and $t^{n-1/2}$.

\begin{figure}
        \centering
        \includegraphics[width=0.95\linewidth]{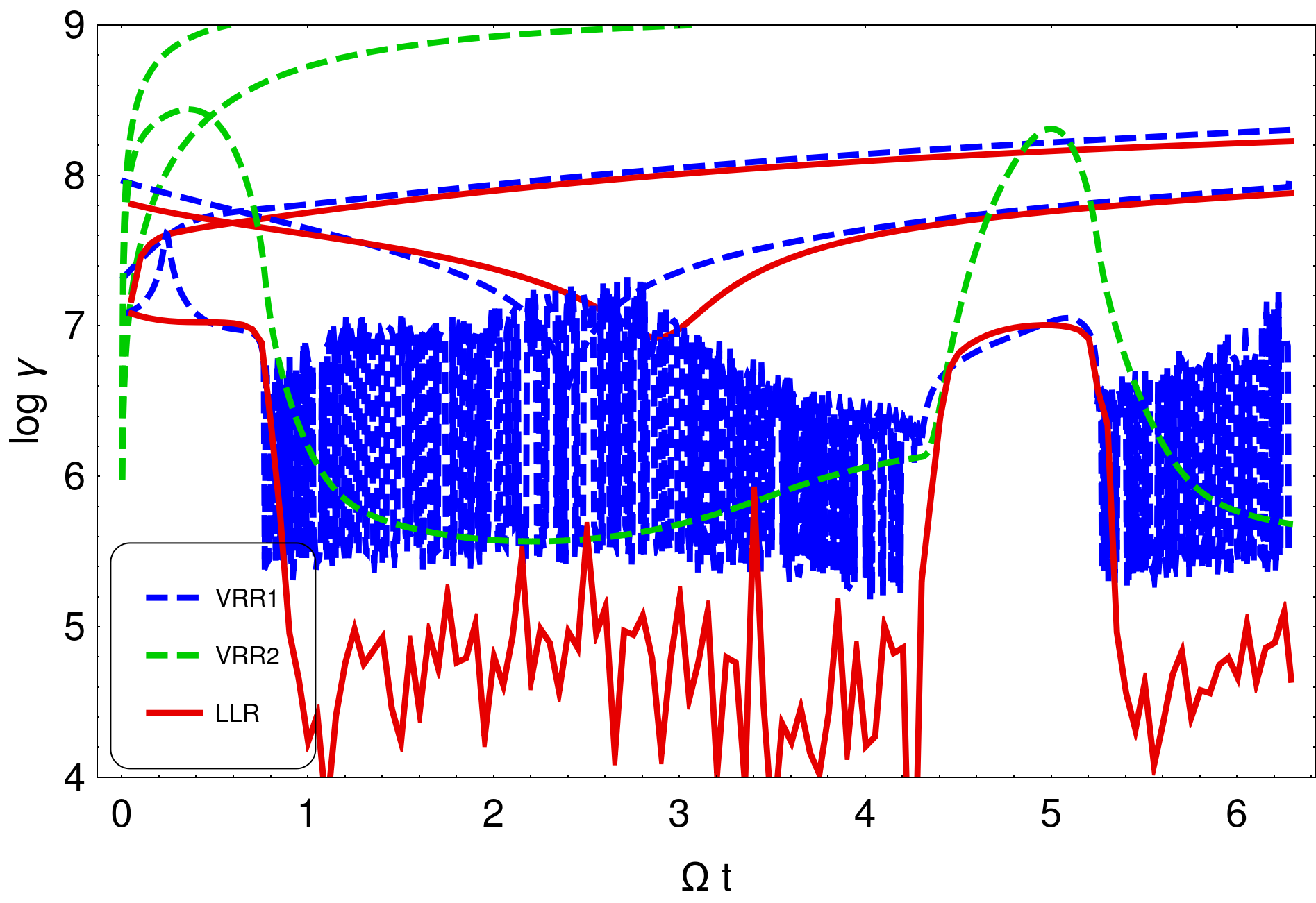}
        \caption{Evolution of the Lorentz factor for the VRR approximation in dashed green and dashed blue lines compared to LLR in solid red lines. VRR2 stands for evaluation by integration in time of the Lorentz factor, whereas VRR1 stands for evaluation of the Lorentz factor by the curvature radius.}
        \label{fig:electronc60a13g6cfl-3vgamma}
\end{figure}

If we integrate the time evolution of the Lorentz factor instead, we get less accurate estimates of the Lorentz factor, as shown in Fig. \ref{fig:electronc60a13g6cfl-3vgamma} in green dashed lines for the VRR approach and indicated as  VRR2. The blue dashed lines show the Lorentz factor computed via the curvature and give similar results to IRR in Fig.~\ref{fig:electronc60a13g6cfl-3gamma}, indicated as VRR1.

\begin{figure*}
  \centering
   \includegraphics[width=0.6\linewidth]{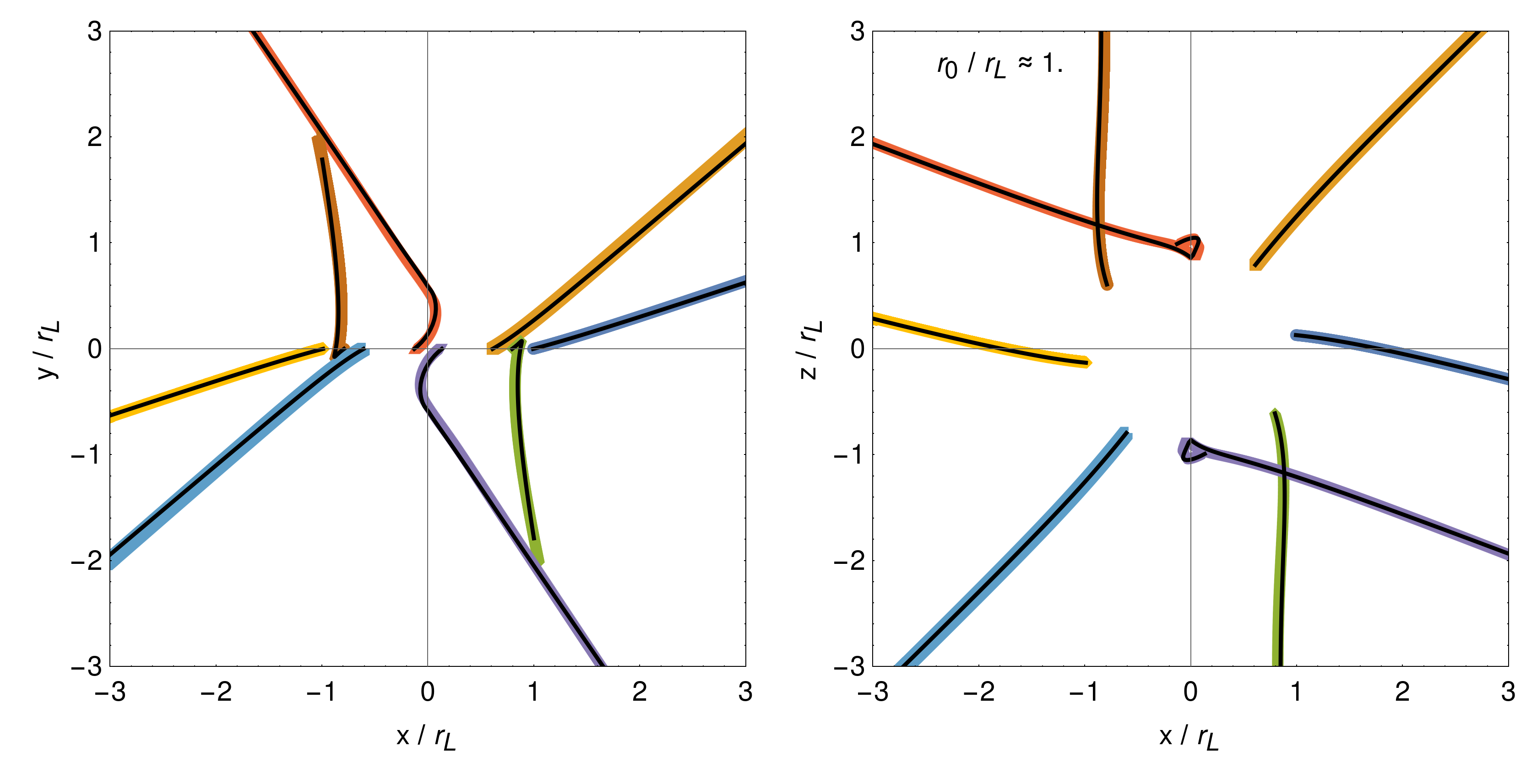} \\
    \vspace{-6mm}
  \includegraphics[width=0.6\linewidth]{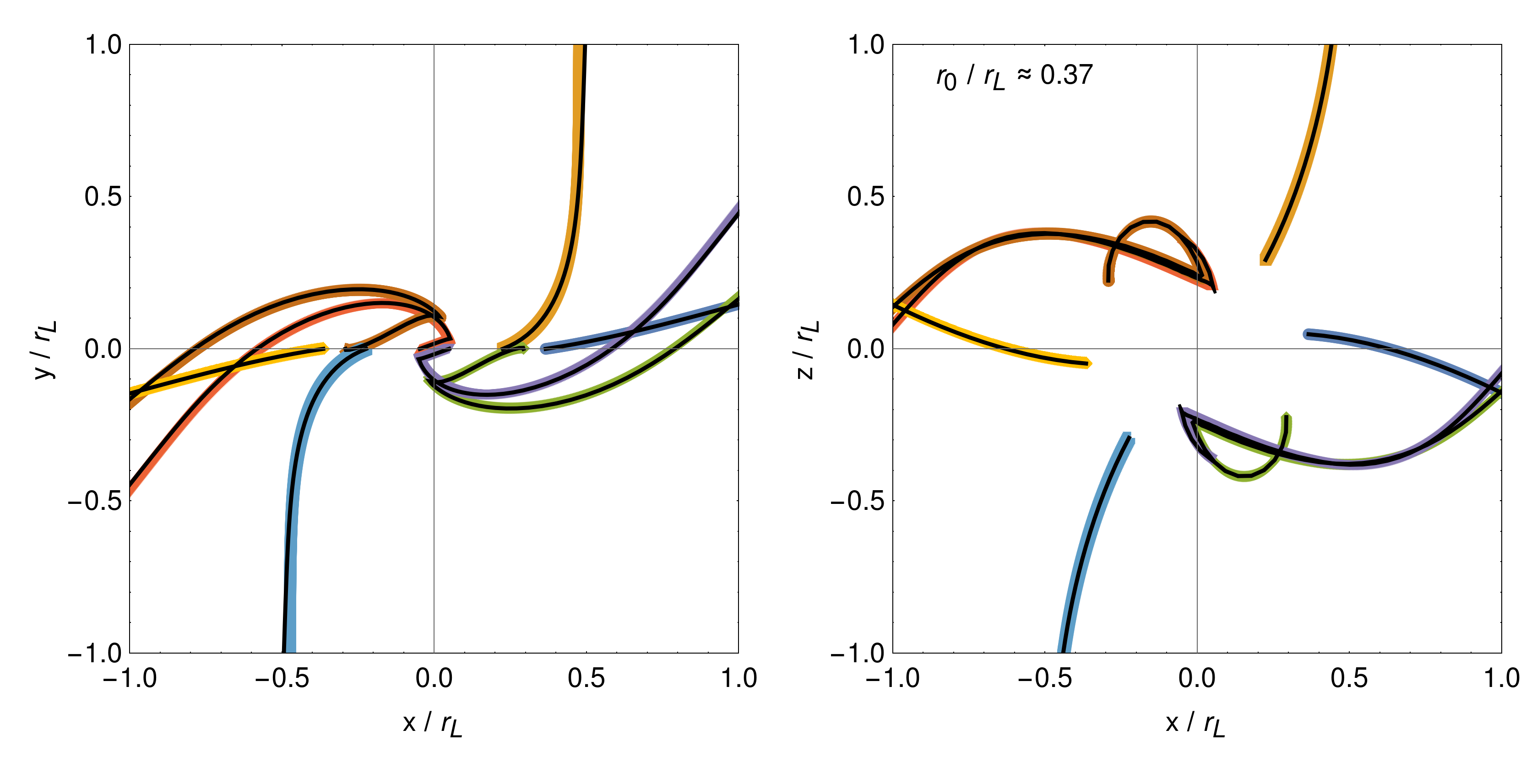} \\
 \vspace{-6mm}
 \includegraphics[width=0.6\linewidth]{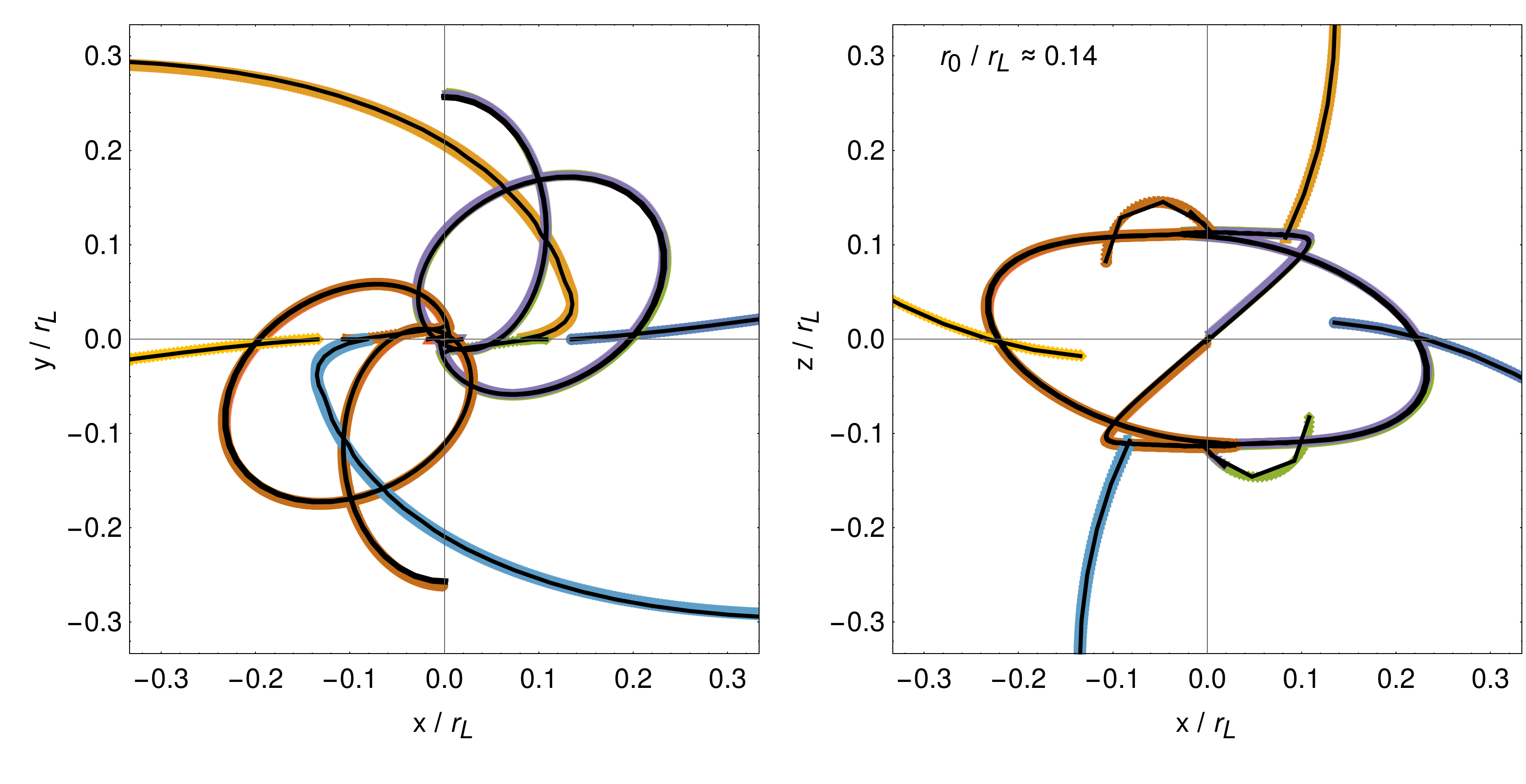} \\
  \vspace{-6mm}
\includegraphics[width=0.6\linewidth]{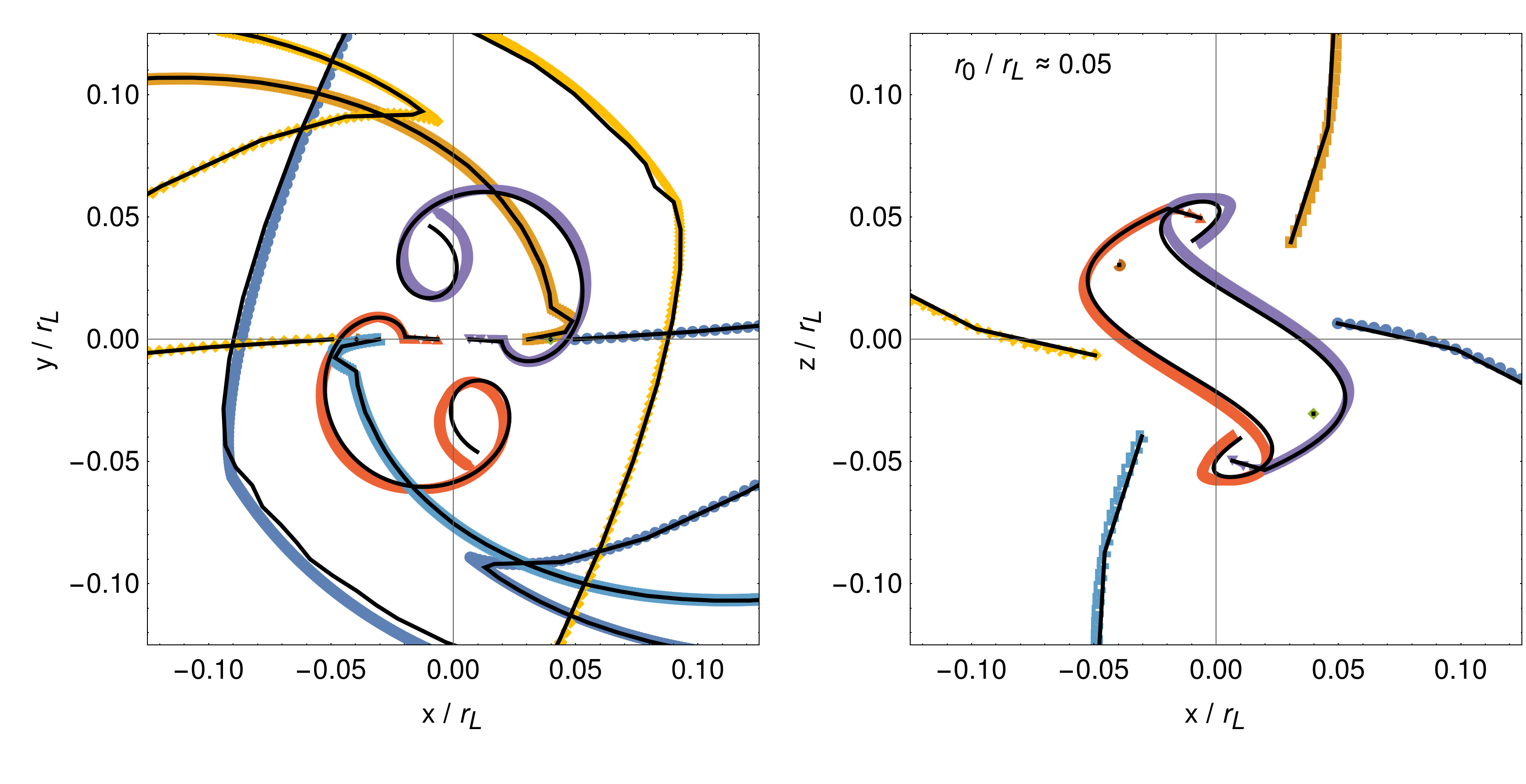}
\caption{Comparison of IRR and LLR trajectories of a relevant sample of electrons. The solid black lines depict the LLR trajectories, and the coloured symbols show the IRR approximation. In each horizontal panel, particles start at a fixed spherical radius given by $r_0/\rlight \approx \{1, 0.37, 0.14, 0.05\}$, from top to bottom. \label{fig:electron_c60_a13_comparaison_ar_llrcfl-3}}
\end{figure*}

Representing another check of the efficiency of the IRR and \RR\ approximation, Fig.~\ref{fig:electron_c60_a13_comparaison_ar_llrcfl-3} overlaps the LLR trajectories shown in black solid lines onto the IRR trajectories shown in coloured, thick solid lines for a sample of particles starting at $r_0/\rlight \approx \{1, 0.37, 0.14, 0.05\}$ from the top to the bottom row, respectively see the legend in the right-column panels). For the trajectories starting well above the stellar surface, corresponding to the first row with $r_0/\rlight \approx 1$ and to the second row with $r_0/\rlight \approx 0.37$, all trajectories agree and overlap. Only some trajectories starting from the stellar surface, corresponding to the fourth row with $r_0/\rlight \approx 0.05,$ do not match, though the behaviour remains the same. For particles starting at $r_0/\rlight \approx 0.14$, third row, the agreement is also excellent.

\begin{figure}
        \centering
        \includegraphics[width=\linewidth]{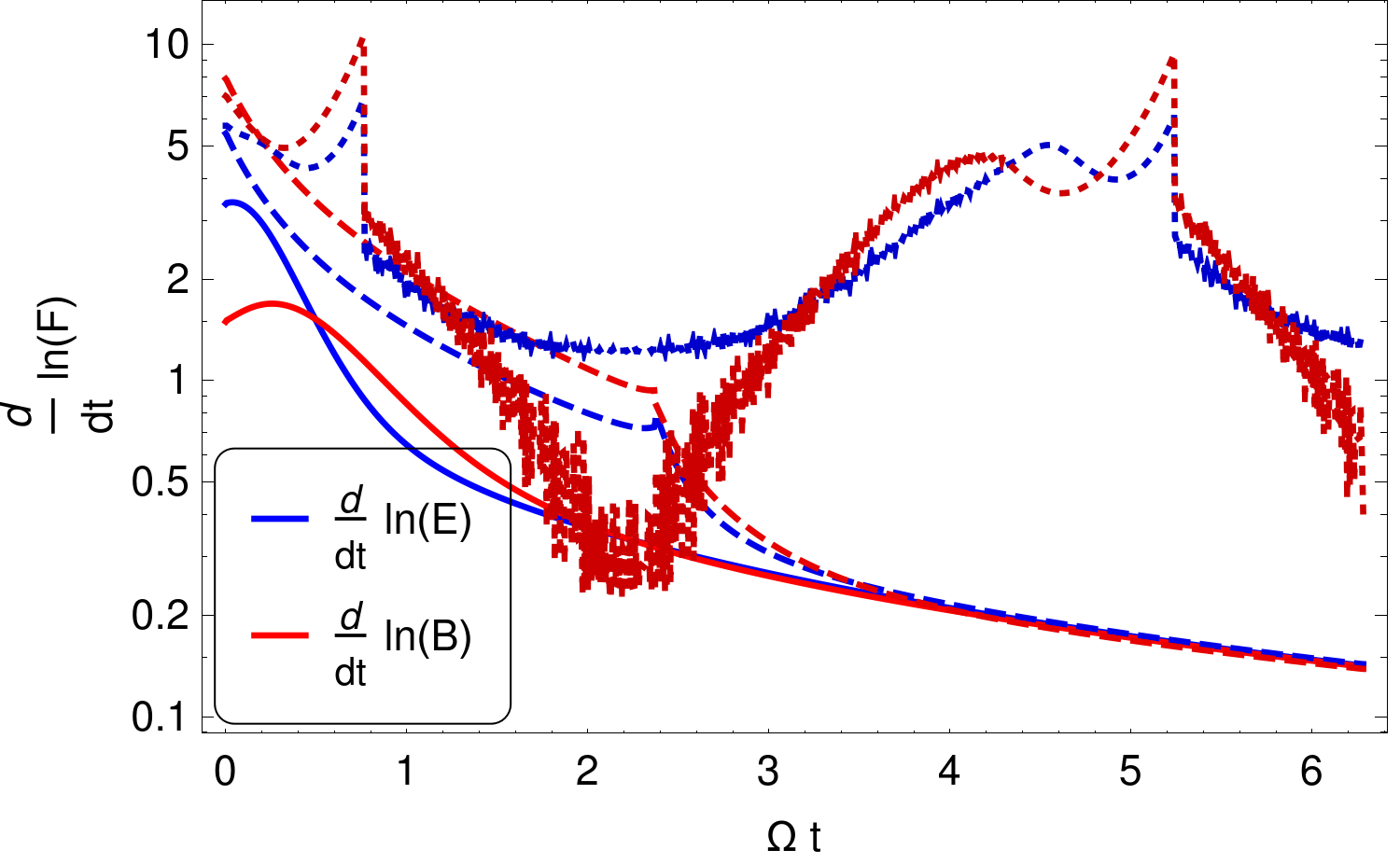}
        \caption{Time evolution of the electric field derivative $d\ln E/dt$ (blue) and magnetic field derivative $d\ln B/dt$ (red), in normalised units, along a sample of three particle trajectories depicted by solid lines, dashed lines, and dotted lines.}
        \label{fig:electronc60a13g6dfsdt}
\end{figure}

Finally, we stress that all the above results rely on the assumption that $d\vec{E}/dt = d\vec{B}/dt = \vec{0}$ are valid. This is correct as long as the terms involving $d\vec{E}/dt$ and $d\vec{B}/dt$ remain small compared to the other terms in Eq.~\eqref{eq:LL3D}. We checked this a posteriori by computing the time derivatives $d\ln E/dt$ and $d\ln B/dt$, in normalised units, during a full simulation span. The time evolution of these derivatives is shown in Fig. \ref{fig:electronc60a13g6dfsdt}. They remain of order unity, with a maximal value of about ten. If multiplied by the correct factors given in the \LL\ equation, we noticed that these terms in the radiation reaction force indeed stay at a negligible level, even when multiplied by a factor $\gamma$. A simple criterion for dropping these terms is $\gamma \, k_2 \ll 1$. Thus, we can confidently ignore these time derivatives even outside the light cylinder.

\section{Conclusions}
\label{sec:Conclusions}

Tracking a charged particle motion in an ultra-strong electromagnetic field is computationally a very demanding task. However, finding accurate approximations able to follow these ultra-relativistic trajectories with radiative friction is a central problem in modelling realistic neutron star magnetospheres. In this paper, we extended the velocity vector expression in the RRL by including a radiative force linear in velocity as derived from the \LL\ equation. We showed that integrating the particle trajectories with this new expression gives very similar results to the `standard' radiation reaction expression of the Aristotelian dynamics. The Lorentz factors are identical in both cases. A new parameter was introduced to control the strength of this force linear in velocity compared to the ultra-relativistic term proportional to $\gamma^2$. It almost always remains negligible compared to the $\gamma^2$ term anti-aligned with the velocity vector. Including such a refinement in the radiation reaction regime to obtain more accurate solutions is therefore not recommended because it also requires more computational time for no benefit.

Nevertheless, we observed some discrepancy between the \LL\ solution and the IRR\ solution for some particles starting from regions close to the surface where the field strength is maximal. In such cases, the \LL\ integration scheme is recommended if accuracy becomes an issue in obtaining reliable results. An alternative approach therefore would be to evolve the velocity vector in time and the Lorentz factor using the ultra-relativistic equation of motion approximation for a charged particle while assuming that the speed is and remains very close to the speed of light.

Another possible application beyond neutron stars but not explored in this work is using lasers in the extreme light regime to investigate high energy physics in ultra-strong electromagnetic fields in the laboratory. Indeed current technology pushes the laser nominal intensity above $I_0 \gtrsim 10^{22}$~W/cm$^2$ \citep{gonoskov_charged_2022}, corresponding to magnetic field strengths on the order of $B\gtrsim10^7$~T, which are similar to field strengths met around compact objects in high energy astrophysics. At such laser intensities, the field strength is expected to reach the radiation dominated regime and even the strong field quantum electrodynamics domain where electron-positron pair cascades are triggered.

\begin{acknowledgements}
I am grateful to the referee for helpful comments and suggestions. This work has been supported by the CEFIPRA grant IFC/F5904-B/2018 and ANR-20-CE31-0010.
\end{acknowledgements}

\end{document}